\documentclass[man,floatsintext,12pt,letterpaper]{apa7}
\usepackage[T1]{fontenc}
\usepackage{lmodern}
\usepackage[american]{babel}
\usepackage{csquotes}
\usepackage[style=apa,backend=biber]{biblatex}
\usepackage{hyperref}
\hypersetup{hidelinks,bookmarksnumbered}
\makeatletter
\let\apasection\section
\renewcommand{\section}{%
  \@ifstar{\apasection*}{%
    \@startsection{section}{1}{\z@}%
      {\b@level@one@skip}{\e@level@one@skip}%
      {\normalfont\normalsize\bfseries\raggedright}}}
\renewcommand{\subsection}{%
  \@startsection{subsection}{2}{\z@}%
    {\b@level@two@skip}{\e@level@two@skip}%
    {\normalfont\normalsize\bfseries\raggedright}}
\newcommand{\section@headingnumber}{\thesection.}
\renewcommand{\@seccntformat}[1]{%
  \@ifundefined{#1@headingnumber}{\csname the#1\endcsname}{\csname #1@headingnumber\endcsname}%
  \enspace}
\makeatother

\usepackage{amsmath,amssymb,amsthm,mathtools}
\usepackage{needspace}
\usepackage{ragged2e}
\usepackage{microtype}
\usepackage{tikz}
\usetikzlibrary{arrows.meta,positioning}

\newcommand{\LRmean}{0.834}

\newcommand{\WidthExample}{0.170}

\DeclareMathOperator{\Cov}{Cov}

\DeclareMathOperator{\tr}{tr}

\newcommand{\dd}{\mathrm{d}}
\newcommand{\E}{\mathbb{E}}
\newcommand{\R}{\mathbb{R}}
\newcommand{\one}{\mathbf{1}}

\title{Analytical Wald Power for Lagged Coefficients in the Random Intercept Cross-Lagged Panel Model}
\shorttitle{ANALYTICAL POWER FOR THE RI-CLPM}
\authorsnames{Zihao Cao}
\authorsaffiliations{East China Normal University}

\authornote{
  \addORCIDlink{Zihao Cao}{0009-0002-0962-7849}

  Correspondence concerning this article should be addressed to Zihao Cao,
  East China Normal University. E-mail: \href{mailto:10180350247@stu.ecnu.edu.cn}{10180350247@stu.ecnu.edu.cn}

  The author used ChatGPT (OpenAI, GPT-6 Astra) to assist with manuscript drafting, language editing, and checking the mathematical derivations and analysis code. The author takes full responsibility for the content of this work.
}

\abstract{\justifying\noindent In a longitudinal study, the number of participants and the number of waves both affect how much power the study has to detect a lagged effect. The random intercept cross-lagged panel model (RI-CLPM) is widely used to study within-person dynamics, but existing approaches to power analysis for this model typically rely on simulation or target likelihood-ratio tests. Starting from the likelihood score, we derive analytical Wald power for each lagged coefficient in the bivariate RI-CLPM. Across 30 designs, we compared our analytical results with power estimates from two R packages, \texttt{powRICLPM} for simulated Wald power and \texttt{semPower} for analytical likelihood-ratio power. The mean absolute difference was less than one percentage point in each comparison. By avoiding repeated simulation and model fitting, the analytical method saves time when evaluating power for a target path across many combinations of sample size (\(N\)) and number of waves (\(T\)).}


\begin{document}
\maketitle
\justifying
\section{Introduction}
\label{sec:introduction}

Psychologists increasingly use longitudinal studies to examine how individuals change over time \parencite{Lafit2021ILPowerShiny}. The random intercept cross-lagged panel model (RI-CLPM) has become a widely used approach to studying reciprocal relations because it separates within-person fluctuations from stable differences between people \parencite{Hamaker2015RICLPM,Mulder2021ThreeExtensions}. Designing such a study requires choosing both the number of participants (\(N\)) and the number of waves (\(T\)). Suppose a study can afford 1,200 assessments. Should it measure 400 people three times or 300 people four times? The first design provides more independent trajectories. The second provides three transitions per person, compared with two in the first. Which \((N,T)\) pair offers greater power depends on the lagged effect the study aims to detect and the dynamics of the two variables.

For many longitudinal models, researchers lack an analytical formula for the power of their chosen test of a specific coefficient \parencite{Hecht2024OptimalNT,Irmer2024MSPE}. Monte Carlo simulation offers a flexible solution. Researchers generate samples under a specified population model, fit the planned analysis model to each sample, and estimate power from the proportion of tests that reject the null hypothesis. This approach has been used for mixed-effects models \parencite{Kumle2021SimPower}, intensive longitudinal regressions with temporal dependence \parencite{Lafit2021ILPowerShiny}, and the bivariate RI-CLPM in the R package \texttt{powRICLPM} \parencite{Mulder2023powRICLPM}. Computational cost increases with the number of candidate designs because comparing many \((N,T)\) pairs requires repeated model fitting at each design. The R package \texttt{mlpwr} uses a surrogate model fitted to simulated power estimates to choose which designs to simulate next, guiding the search for a cost-efficient design that reaches the desired power \parencite{Zimmer2024mlpwr}.

For some models and design criteria, power can be calculated directly from population parameters. In dynamic panel models including the RI-CLPM, \textcite{Hecht2024OptimalNT} used analytical likelihood-ratio (LR) power to choose the numbers of participants and waves under a fixed budget. The R package \texttt{semPower} also evaluates LR power for hypotheses expressed as restrictions on a structural equation model \parencite{MoshagenBader2024semPower}. For a univariate random intercept panel, \textcite{Usami2025OptimalDesign} used the standard error of the autoregressive coefficient to compare designs by confidence interval width. Usami identified the multivariate RI-CLPM as a next step. The bivariate RI-CLPM has a more complex covariance structure because it must also account for associations between the two variables, both within and across waves. To our knowledge, no published study has derived an analytical expression for the Wald power of individual lagged coefficients in this model.

We derive an analytical expression for the Wald power of each autoregressive and cross-lagged coefficient in the bivariate RI-CLPM. For each path, the calculation also gives the design standard error and corresponding confidence interval width. Across 30 designs, the mean absolute difference was less than one percentage point in each comparison, whether we used simulated Wald power from the R package \texttt{powRICLPM} or analytical LR power from the R package \texttt{semPower}. The analytical calculation saves the time needed for repeated simulation, making it practical to evaluate a target path across many candidate \((N,T)\) pairs.

\section{The Random Intercept Cross-Lagged Panel Model}
\label{sec:model}

\subsection{Observations and Latent Components}

Consider a panel of \(N\) independent individuals observed over \(T\ge 3\) waves. At least three waves are necessary to identify the 13 covariance-structure parameters \parencite{Hamaker2015RICLPM}. Let \(y_{it}\in\R^2\) denote the observed pair for individual \(i\) at wave \(t\). We assume that both variables are measured for everyone at every wave. The same four lagged coefficients link each wave to the next. The first equation below shows what makes up each observed score. The second shows how a person's deviations from their usual levels carry over to the next wave.
\begin{align}
 y_{it}&=\mu_t+\eta_i+w_{it}, &&t=1,\ldots,T, \label{eq:observation}\\
 w_{it}&=F w_{i,t-1}+\varepsilon_{it}, &&t=2,\ldots,T, \label{eq:within}\\
 F&=\begin{pmatrix}a&b\\c&d\end{pmatrix}. \nonumber
\end{align}
The wave means \(\mu_t\) may be treated as known or estimated freely. The first and second diagonal entries of \(F\) are the autoregressive coefficients for the first and second variables. Entries \(b\) and \(c\) are the paths from the second variable to the first and from the first to the second, respectively. This indexing applies to raw-scale within-person scores.

The random intercept \(\eta_i\), the initial state \(w_{i1}\), and innovations \(\varepsilon_{i2},\ldots,\varepsilon_{iT}\) are mutually independent, with
\begin{equation}
 \eta_i\sim\mathcal N_2(0,\Omega),\qquad
 w_{i1}\sim\mathcal N_2(0,\Gamma),\qquad
 \varepsilon_{it}\sim\mathcal N_2(0,\Psi),\qquad
 \Omega,\Gamma,\Psi\succ0 .
 \label{eq:inputs}
\end{equation}
At the first wave, \(\Gamma\) describes how much the two within-person deviations vary and how they vary together. At later waves, \(\Psi\) describes the variation that remains after accounting for the previous wave. To generate the simulated data, we chose values satisfying \(\Gamma=F\Gamma F^\top+\Psi\) so that the two within-person variances and their covariance stayed the same across waves. When fitting the model, we estimated the two first-wave variances and their covariance freely, without imposing this equality.

\subsection{Stacked Covariance and Block Structure}

Stack the within-person states in chronological order and write \(K=\Cov(w_i,w_i)\). Each block \(K_{ts}=\Cov(w_{it},w_{is})\) is \(2\times2\). Start with \(K_{11}=\Gamma\). At wave \(t\), the previous state enters both sides of its covariance and the new innovation is independent of the past, so
\begin{equation}
 K_{tt}=FK_{t-1,t-1}F^\top+\Psi,\qquad t\ge2 .
 \label{eq:diagonal-K}
\end{equation}
For a pair \(t>s\), only the later state needs to be propagated one step. Taking its covariance with \(w_{is}\) gives
\begin{equation}
 K_{ts}=F K_{t-1,s},\qquad K_{st}=K_{ts}^\top .
 \label{eq:off-diagonal-K}
\end{equation}
We can therefore compute the diagonal blocks first. Column \(s\) below the diagonal follows by repeated left multiplication of \(K_{ss}\) by \(F\). For example, \(K_{21}=F\Gamma\), \(K_{31}=F K_{21}\), and \(K_{32}=F K_{22}\).

Every pair of observed waves shares the same random intercept. Since \(\eta_i\) is independent of the within-person process, \(\Sigma_{ts}=K_{ts}+\Omega\). With \(\one_T\) the length-\(T\) vector of ones, the full covariance and a useful factorization of its random intercept term are
\begin{align}
 \Sigma&=K+(\one_T\one_T^\top)\otimes\Omega,\label{eq:Sigma}\\
 (\one_T\one_T^\top)\otimes\Omega
   &=(\one_T\otimes I_2)\,\Omega\,(\one_T^\top\otimes I_2).
 \label{eq:kron}
\end{align}
The second line follows from the mixed-product rule for Kronecker products, treating \(\Omega\) as \([1]\otimes\Omega\).

\subsection{Parameters and Differential Convention}

We order the 13 covariance-structure parameters as
\begin{equation}
 \theta=(a,b,c,d,\sigma_x^2,\sigma_{xy},\sigma_y^2,
 \kappa_x,\kappa_{xy},\kappa_y,\tau_x,\tau_{xy},\tau_y)^\top ,
 \label{eq:theta}
\end{equation}
where \((\sigma_x^2,\sigma_{xy},\sigma_y^2)\), \((\kappa_x,\kappa_{xy},\kappa_y)\), and \((\tau_x,\tau_{xy},\tau_y)\) are the two variances and covariance of \(\Psi,\Gamma,\Omega\), respectively. Each variance is a separate parameter. When we estimate the wave means freely, we treat them as a separate set of parameters.

Following the matrix differential calculus of \textcite{MagnusNeudecker2019}, we write \(\dd M\) for the total differential of a matrix-valued function \(M(\theta)\). To obtain \(\partial M/\partial\theta_j\), we hold all other parameters fixed and read off the matrix multiplying \(\dd\theta_j\). For example, with \(E_{uv}\) the \(2\times2\) matrix unit, perturbing \(b\) gives \(\dd F=E_{12}\,\dd b\). Varying the covariance \(M_{12}=M_{21}\) gives \(\dd M=(E_{12}+E_{21})\,\dd M_{12}\) in its associated symmetric covariance matrix \(M\).

\section{Covariance Derivatives and Structured Precision}
\label{sec:precision}

\subsection{Recursion for the Covariance Derivatives}

At fixed \(T\), the model equations define a function from the 13 population parameters to the trajectory covariance, \(\theta\mapsto\Sigma(\theta;T)\). Its partial derivative \(D_j(\theta_0;T)\) has the same dimensions as \(\Sigma\). Entry \((r,s)\) gives the rate of change of covariance \(\Sigma_{rs}\) with respect to \(\theta_j\) at the specified population value \(\theta_0\), holding the other 12 parameters fixed. The algorithm calculates this rate directly. It differentiates the covariance recursions by the product rule and evaluates the resulting formulas at \(\theta_0\).

Differentiating \eqref{eq:Sigma} with fixed \(T\) gives
\begin{equation}
 \dd\Sigma=\dd K+(\one_T\one_T^\top)\otimes\dd\Omega.
 \label{eq:dSigma}
\end{equation}
The Kronecker factor \(\one_T\one_T^\top\) is fixed by the observation design. Differentiating \eqref{eq:diagonal-K} and \eqref{eq:off-diagonal-K} yields the block updates
\begin{align}
 \dd K_{11}&=\dd\Gamma, \nonumber\\
 \dd K_{tt}
 &= (\dd F)K_{t-1,t-1}F^\top+F(\dd K_{t-1,t-1})F^\top \nonumber\\
 &\quad +F K_{t-1,t-1}(\dd F)^\top+\dd\Psi,\qquad t\ge2,\\
 \dd K_{ts}&=(\dd F)K_{t-1,s}+F\,\dd K_{t-1,s},\quad
 \dd K_{st}=(\dd K_{ts})^\top,\qquad t>s .
 \label{eq:dK}
\end{align}
The middle term carries the change in the previous wave's covariance into the current wave. The other terms account for changes in \(F\) on either side of the covariance matrix and in \(\Psi\).

To obtain \(D_j\), replace each differential by its coefficient of \(\dd\theta_j\). For the four entries of \(F\), the directions are \(E_{11},E_{12},E_{21},E_{22}\). For each symmetric covariance matrix the three directions are \(E_{11},E_{12}+E_{21},E_{22}\). For example, \(b=F_{12}\) appears in exactly one entry of the transition matrix, so
\[
 \frac{\partial F}{\partial b}=E_{12}
 =\begin{pmatrix}0&1\\0&0\end{pmatrix},\qquad
 \frac{\partial\Gamma}{\partial b}
 =\frac{\partial\Psi}{\partial b}
 =\frac{\partial\Omega}{\partial b}=0_{2\times2}.
\]
The entry 1 is the derivative of \(F_{12}=b\) with respect to itself. The algorithm first computes the covariance blocks \(K_{ts}\) at \(\theta_0\). It then uses the four input derivatives above in \eqref{eq:dK}, starting with \(\partial K_{11}/\partial b=0\). At wave 2, differentiating \(\Sigma_{21}=F\Gamma+\Omega\) and \(\Sigma_{22}=F\Gamma F^\top+\Psi+\Omega\) gives
\begin{align}
 (D_b)_{21}&=E_{12}\Gamma
   =\begin{pmatrix}\kappa_{xy}&\kappa_y\\0&0\end{pmatrix}, \nonumber\\
 (D_b)_{22}&=E_{12}\Gamma F^\top+F\Gamma E_{12}^\top.
 \label{eq:Db-example}
\end{align}
Every matrix on the right is evaluated at \(\theta_0\). In \((D_b)_{22}\), one term differentiates the left occurrence of \(F\) and the other differentiates \(F^\top\). The same product rule propagates these derivatives to later waves. This calculation uses only the parameter values at \(\theta_0\) and requires no finite-difference step size. A change in \(\tau_{xy}\) instead adds \(E_{12}+E_{21}\) to the derivative of every observed block. Applying the recursions to the 13 directions produces the full set of symmetric \(2T\times2T\) matrices \(D_j\).

\subsection{Transforming States into Independent Inputs}

Subtracting \(F w_{i,t-1}\) from \(w_{it}\) recovers the innovation at wave \(t\). For three waves this transformation is
\begin{equation}
 \underbrace{\begin{pmatrix}I_2&0&0\\-F&I_2&0\\0&-F&I_2\end{pmatrix}}_{\mathcal L}
 \begin{pmatrix}w_{i1}\\w_{i2}\\w_{i3}\end{pmatrix}
 =\begin{pmatrix}w_{i1}\\\varepsilon_{i2}\\\varepsilon_{i3}\end{pmatrix}.
 \label{eq:L}
\end{equation}
For general \(T\), repeat the same diagonal and subdiagonal blocks. Independence of these inputs gives
\begin{equation}
 \mathcal L K\mathcal L^\top=\mathcal D,\qquad
 \mathcal D=\operatorname{blockdiag}(\Gamma,\Psi,\ldots,\Psi).
 \label{eq:LK}
\end{equation}
The unit diagonal makes \(\mathcal L\) invertible. Solving for \(K\) and reversing the order of factors upon inversion yields
\begin{equation}
 Q=K^{-1}=\mathcal L^\top\mathcal D^{-1}\mathcal L.
 \label{eq:Q}
\end{equation}
The independence and positive definiteness in \eqref{eq:inputs} imply \(K\succ0\); adding the positive-semidefinite random intercept term in \eqref{eq:Sigma} ensures \(\Sigma\succ0\).

\subsection{Block-Tridiagonal Within-Person Precision}

The \((t,s)\) block in \eqref{eq:Q} is
\begin{equation}
 Q_{ts}=\sum_{r=1}^T \mathcal L_{rt}^\top
          (\mathcal D^{-1})_{rr}\mathcal L_{rs}.
 \label{eq:Q-block}
\end{equation}
Column \(t\) of \(\mathcal L\) has nonzero blocks only at rows \(t\) and \(t+1\), so columns \(t\) and \(s\) have no common contributing block row when \(|t-s|>1\). For example, only row 2 contributes to \(Q_{21}\), giving \(I_2\Psi^{-1}(-F)=-\Psi^{-1}F\). Both the initial state and the second-wave innovation contribute to the first diagonal block. Only the final innovation contributes to the last diagonal block. Writing \(G=\Gamma^{-1}\) and \(R=\Psi^{-1}\), the four-wave pattern is
\begin{equation}
 Q=\begin{pmatrix}
 G+F^\top RF&-F^\top R&0&0\\
 -RF&R+F^\top RF&-F^\top R&0\\
 0&-RF&R+F^\top RF&-F^\top R\\
 0&0&-RF&R
 \end{pmatrix}.
 \label{eq:Q4}
\end{equation}
The middle block rows repeat at additional waves.

\subsection{Inverting the Covariance Matrix with Random Intercepts}

Apply the Woodbury identity to the factored random intercept term in \eqref{eq:kron}. The central matrix in the correction is
\begin{equation}
 C=\Omega^{-1}+(\one_T^\top\otimes I_2)Q(\one_T\otimes I_2).
 \label{eq:C-def}
\end{equation}
The second term maps from two random intercept coordinates to \(2T\) observed coordinates and back to two. Let \(A=I_2-F\). In \(\mathcal L(\one_T\otimes I_2)\), the first block row is \(I_2\) and the remaining \(T-1\) block rows are all \(A\). Substitution of \eqref{eq:Q} then gives
\begin{equation}
 C=\Omega^{-1}+G+(T-1)A^\top R A\succ0 .
 \label{eq:C-explicit}
\end{equation}
To read the outer term, let \(V_t\) be block \(t\) of \(Q(\one_T\otimes I_2)\). Multiplying first by \(\mathcal D^{-1}\mathcal L\) produces the blocks \(G,RA,\ldots,RA\). The final multiplication by \(\mathcal L^\top\) subtracts \(F^\top\) times the next block, if there is one. Consequently,
\begin{equation}
 V_t=\begin{cases}
 G-F^\top RA,&t=1,\\
 A^\top RA,&2\le t<T,\\
 RA,&t=T.
 \end{cases}
 \qquad
 \boxed{P_{ts}=Q_{ts}-V_tC^{-1}V_s^\top.}
 \label{eq:P}
\end{equation}
The right outer factor is the transpose of the left one because \(Q\) is symmetric. Equations \eqref{eq:Q4} and \eqref{eq:P} give the exact observed precision using inverses of dimension two. The random intercept can contribute a nonzero correction even for nonadjacent waves.

\section{Fisher Information and Analytical Wald Power}
\label{sec:information}

\subsection{The Gaussian Score and Expected Information}

Let \(z_i=y_i-\mu\) be the centered \(2T\)-dimensional observation vector. Up to an additive constant, the single-person log-likelihood is
\begin{equation}
 \ell_i(\theta)=-\tfrac12\log\det\Sigma-\tfrac12z_i^\top Pz_i .
 \label{eq:lik}
\end{equation}
For an observed trajectory, \(z_i\) is held fixed while taking covariance derivatives. The log-determinant differential is \(\dd\log\det\Sigma=\tr(P\,\dd\Sigma)\). Differentiating \(P\Sigma=I_{2T}\) and multiplying on the right by \(P\) gives \(\dd P=-P(\dd\Sigma)P\). Thus
\begin{equation}
 \dd\ell_i=-\tfrac12\tr(P\,\dd\Sigma)
          +\tfrac12 z_i^\top P(\dd\Sigma)Pz_i .
 \label{eq:dlik}
\end{equation}
In the direction \(\theta_j\), substituting \(\dd\Sigma=D_j\,\dd\theta_j\) and reading the coefficient of \(\dd\theta_j\) yields the score
\begin{equation}
 u_{ij}=\frac{\partial\ell_i}{\partial\theta_j}
   =\tfrac12\{z_i^\top P D_jPz_i-\tr(PD_j)\}.
 \label{eq:score}
\end{equation}

The expected information is defined as the expected outer product of the score vector \(u_i\), so \([\mathcal I_1]_{jk}=\E_\theta[u_{ij}u_{ik}]\). Write \(A_j=P D_jP\) and \(q_j=z_i^\top A_jz_i\). Using \(\E_\theta(z_i z_i^\top)=\Sigma\) and taking the trace before or after the expectation gives \(\E_\theta(q_j)=\tr(A_j\Sigma)=\tr(PD_j)\). Each score in \eqref{eq:score} is therefore a centered quadratic form divided by two. For symmetric \(A_j,A_k\), the Gaussian quadratic-form covariance is \(2\tr(A_j\Sigma A_k\Sigma)\). Hence
\begin{equation}
 \begin{aligned}
 [\mathcal I_1]_{jk}
  &=\tfrac14\Cov(q_j,q_k)
   =\tfrac12\tr\{(P D_j P)\Sigma(P D_k P)\Sigma\}\\
  &=\tfrac12\tr(P D_j P D_k).
 \end{aligned}
 \label{eq:fisher}
\end{equation}
The last equality uses \(P\Sigma=I_{2T}\) twice while preserving the order of the factors. When wave means are free, the mean score is linear in the centered observations and the covariance score is a centered quadratic form. Gaussian third central moments are zero, so estimating those means has zero expected cross-information with the 13 covariance-structure parameters.

\subsection{Variance Factors Under Joint Estimation}

Assume henceforth that \(\mathcal I_1\succ0\). The total score is the sum of \(N\) independent, mean-zero single-person scores, so \(\mathcal I_N=N\mathcal I_1\). To see the role of joint estimation, write \(\theta=(\beta^\top,\xi^\top)^\top\), with \(\beta=(a,b,c,d)^\top\) and \(\xi\) the nine covariance parameters. For the first block column \((X^\top,Z^\top)^\top\) of \(\mathcal I_1^{-1}\), multiplication by \(\mathcal I_1\) gives
\begin{equation}
 \begin{pmatrix}I_{\beta\beta}&I_{\beta\xi}\\
 I_{\xi\beta}&I_{\xi\xi}\end{pmatrix}
 \begin{pmatrix}X\\Z\end{pmatrix}=\begin{pmatrix}I_4\\0\end{pmatrix}.
 \label{eq:block-inverse}
\end{equation}
The lower row implies \(Z=-I_{\xi\xi}^{-1}I_{\xi\beta}X\), because \(I_{\xi\xi}\) is a positive-definite principal submatrix. Substituting into the upper row produces
\begin{equation}
 X=S^{-1},\qquad S=I_{\beta\beta}-I_{\beta\xi}I_{\xi\xi}^{-1}I_{\xi\beta}.
 \label{eq:schur}
\end{equation}
Thus the covariance parameters enter through the Schur correction and the four slopes still interact through inversion of \(S\). For numerical evaluation, one may invert the full \(13\times13\) information matrix and extract its first four diagonal entries, yielding the same \(f_{jj}=[\mathcal I_1^{-1}]_{jj}\).

\subsection{From the Maximum-Likelihood Limit to Wald Power}

For a consistent interior maximum-likelihood estimator in a regular, correctly specified model at fixed \(T\), the total score evaluated at the true parameter and divided by \(\sqrt N\) converges to \(\mathcal N_{13}(0,\mathcal I_1)\). The negative Hessian divided by \(N\) converges in probability to \(\mathcal I_1\). Taylor expansion of the score along the segment from \(\theta\) to \(\widehat\theta\), followed by Slutsky's theorem \parencite{VDV1998}, gives
\begin{equation}
 \sqrt N(\widehat\theta-\theta)\xrightarrow{d}
 \mathcal N_{13}(0,\mathcal I_1^{-1}).
 \label{eq:MLE}
\end{equation}
For coefficient \(j\) the design approximation is \(\widehat\theta_j\;\dot{\sim}\;\mathcal N(\theta_j,f_{jj}/N)\). The variance in this large-sample approximation comes from the exact information for the chosen number of waves.

Consider \(H_0:\theta_j=\beta_{j,0}\), with \(z_*=\Phi^{-1}(1-\alpha/2)\) and \(\Phi\) the standard normal distribution function. With a consistent standard error estimate, the Wald statistic converges to a standard normal distribution under the null hypothesis. The two-sided rule rejects outside \([-z_*,z_*]\). For planning, fix the alternative population and evaluate \(s_j=\sqrt{f_{jj}(\theta,T)/N}\) at that point. The planning statistic decomposes as
\begin{equation}
 \frac{\widehat\theta_j-\beta_{j,0}}{s_j}
 =\frac{\widehat\theta_j-\theta_j}{s_j}
   +\underbrace{\frac{\theta_j-\beta_{j,0}}{s_j}}_{\delta_j}.
 \label{eq:wald-shift}
\end{equation}
The first term is approximately standard normal. Evaluating both tails of the shifted distribution at the same null critical values yields
\begin{equation}
 \boxed{\displaystyle
 \pi_j^{\mathrm W}(\theta,N,T,\alpha)
 =\Phi(\delta_j-z_*)+\Phi(-\delta_j-z_*),\qquad
 \delta_j=(\theta_j-\beta_{j,0})\sqrt{\frac{N}{f_{jj}(\theta,T)}} .}
 \label{eq:wald-power}
\end{equation}
Here \(\pi_j^{\mathrm W}\) denotes the analytical power approximation. Under regular local alternatives \(\theta_N=\theta^{(0)}+h/\sqrt N\) with \(\theta_j^{(0)}=\beta_{j,0}\), the actual Wald statistic has limit \(\mathcal N(h_j/\sqrt{f_{jj}(\theta^{(0)},T)},1)\). At a fixed alternative and finite \(N\), \eqref{eq:wald-power} uses the population standard error as a planning approximation to the estimated standard error.

The same standard error gives the design full width of a two-sided \(100(1-\alpha)\%\) Wald interval,
\begin{equation}
 W_j=2z_*s_j=2z_*\sqrt{f_{jj}(\theta,T)/N}.
 \label{eq:width}
\end{equation}
After data collection, the interval width varies from sample to sample because it uses the estimated standard error. A study seeking a high probability of achieving a specified width would also need the sampling distribution of that width.

\begin{figure}[H]
\centering
\caption{From Population Values Through Analytical Covariance Derivatives to Wald Power}
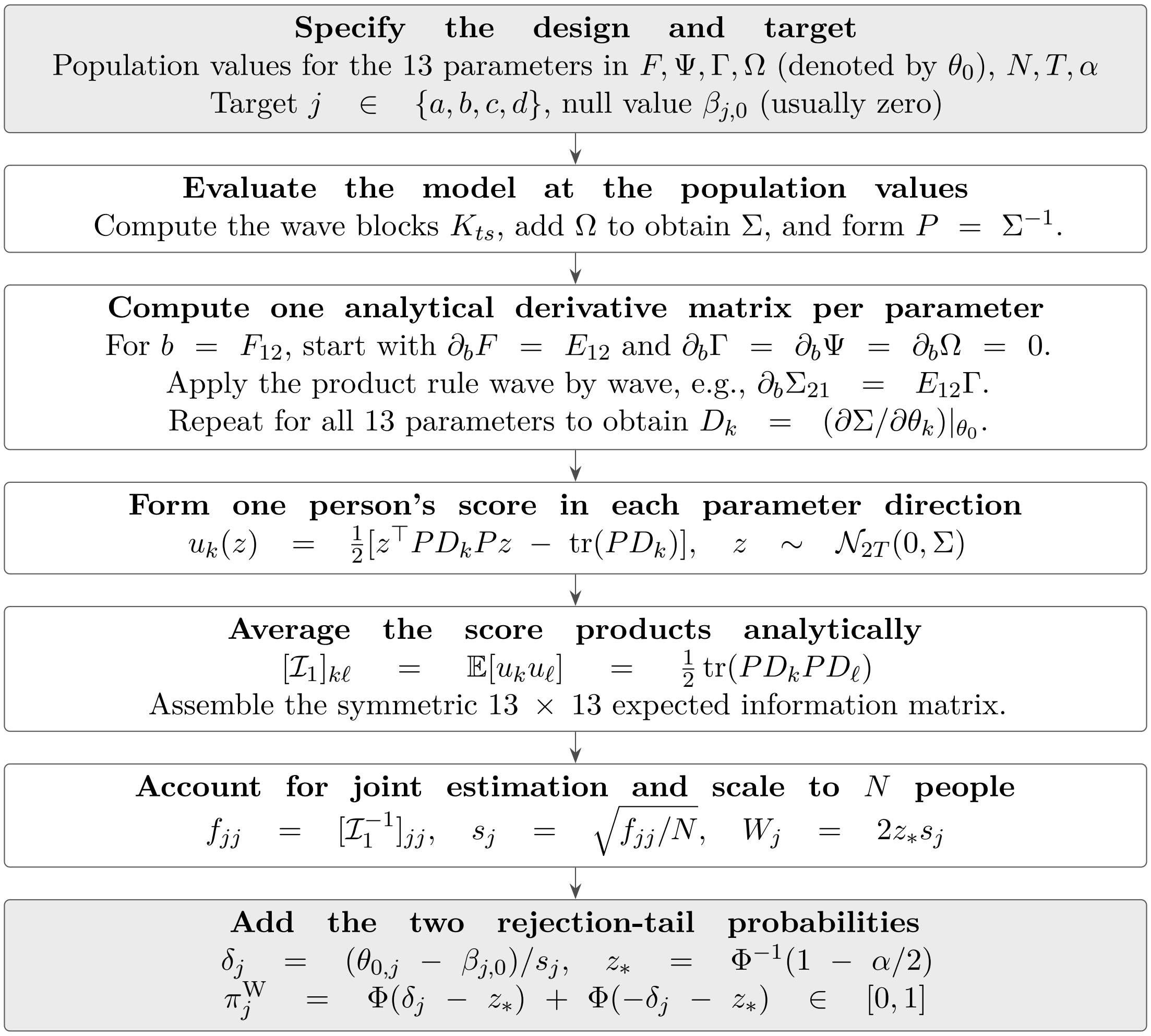
\label{fig:score-to-power}
\begin{tikzpicture}[
 box/.style={draw=black!65,rounded corners=2pt,align=center,
   text width=.86\linewidth,inner sep=4pt,font=\small\linespread{1}\selectfont},
 flow/.style={-{Stealth[length=2mm]},draw=black!70},
 node distance=4mm]
\node[box,fill=black!8] (input) {\textbf{Specify the design and target}\\
 Population values for the 13 parameters in \(F,\Psi,\Gamma,\Omega\) (denoted by \(\theta_0\)),
 \(N,T,\alpha\)\\
 Target \(j\in\{a,b,c,d\}\), null value \(\beta_{j,0}\) (usually zero)};
\node[box,below=of input] (mom) {\textbf{Evaluate the model at the population values}\\
 Compute the wave blocks \(K_{ts}\), add \(\Omega\) to obtain \(\Sigma\), and form \(P=\Sigma^{-1}\).};
\node[box,below=of mom] (deriv) {\textbf{Compute one analytical derivative matrix per parameter}\\
 For \(b=F_{12}\), start with \(\partial_b F=E_{12}\) and
 \(\partial_b\Gamma=\partial_b\Psi=\partial_b\Omega=0\).\\
 Apply the product rule wave by wave, e.g., \(\partial_b\Sigma_{21}=E_{12}\Gamma\).\\
 Repeat for all 13 parameters to obtain \(D_k=(\partial\Sigma/\partial\theta_k)|_{\theta_0}\).};
\node[box,below=of deriv] (score) {\textbf{Form one person's score in each parameter direction}\\
 \(u_k(z)=\tfrac12[z^\top P D_k Pz-\tr(PD_k)],\quad z\sim\mathcal N_{2T}(0,\Sigma)\)};
\node[box,below=of score] (info) {\textbf{Average the score products analytically}\\
 \([\mathcal I_1]_{k\ell}=\E[u_k u_\ell]=\tfrac12\tr(PD_kPD_\ell)\)\\
 Assemble the symmetric \(13\times13\) expected information matrix.};
\node[box,below=of info] (se) {\textbf{Account for joint estimation and scale to \(N\) people}\\
 \(f_{jj}=[\mathcal I_1^{-1}]_{jj},\quad s_j=\sqrt{f_{jj}/N},\quad W_j=2z_*s_j\)};
\node[box,fill=black!8,below=of se] (out) {\textbf{Add the two rejection-tail probabilities}\\
 \(\delta_j=(\theta_{0,j}-\beta_{j,0})/s_j,\quad z_*=\Phi^{-1}(1-\alpha/2)\)\\
 \(\pi_j^{\mathrm W}=\Phi(\delta_j-z_*)+\Phi(-\delta_j-z_*)\in[0,1]\)};
\foreach \a/\b in {input/mom,mom/deriv,deriv/score,score/info,info/se,se/out}
  \draw[flow] (\a.south) -- (\b.north);
\end{tikzpicture}
\par\smallskip{\footnotesize\textit{Note.} The derivative step evaluates the product-rule formulas in Section~\ref{sec:precision} at the specified parameter values \(\theta_0\).\par}
\end{figure}

\Needspace{0.78\textheight}
Figure~\ref{fig:score-to-power} traces both design outputs to the same score.

\section{Numerical Evaluation and Power Comparisons}
\label{sec:verification}

\subsection{A Four-Wave Worked Example}

Suppose a study measures 400 participants at four waves. For each lagged coefficient, we calculate the power of a two-sided Wald test of the null hypothesis that the coefficient is zero, at \(\alpha=.05\). For this example, we assume the following population values for the lagged coefficients in \(F\) and the covariance matrices \(\Psi\), \(\Gamma\), and \(\Omega\).
\begin{equation}
 \begin{aligned}
 F&=\begin{pmatrix}.30&.08\\.16&.30\end{pmatrix},
 &\Psi&=\begin{pmatrix}.8892&.19716\\.19716&.8556\end{pmatrix},\\
 \Gamma&=\begin{pmatrix}1&.30\\.30&1\end{pmatrix},
 &\Omega&=\begin{pmatrix}3/7&9/70\\9/70&3/7\end{pmatrix}.
 \end{aligned}
 \label{eq:example-population}
\end{equation}
First, construct the \(8\times8\) covariance from its \(2\times2\) blocks. For instance, \(K_{21}=F\Gamma=\bigl(\begin{smallmatrix}.324&.170\\.250&.348\end{smallmatrix}\bigr)\), and \(\Sigma_{21}=K_{21}+\Omega\). Compute the observed precision, either from \eqref{eq:P} or by numerical factorization. Our R code uses Cholesky factorization for this step. Evaluate the differential recursions at the parameter values in \eqref{eq:example-population}. For \(b\), the first cross-wave derivative is \((D_b)_{21}=\bigl(\begin{smallmatrix}.30&1\\0&0\end{smallmatrix}\bigr)\).

Second, multiply \(P\) by each derivative and take the traces in \eqref{eq:fisher}. For instance, \(\tfrac12\tr((PD_b)^2)=2.438105\). For an off-diagonal entry, the two derivatives in the product correspond to different parameters. Third, invert the entire \(13\times13\) information matrix, and read its first four diagonal entries. For \(b\), this yields \(f_{bb}=0.75072009\), \(s_b=0.04332205\), and \(\delta_b=0.08/s_b\approx1.846635\). The sum of the two tails in \eqref{eq:wald-power} is approximately \(0.45495515\). Table~\ref{tab:worked-example} gives all four outputs and the corresponding Monte Carlo results for this design. For these parameter values, the design full width of the 95\% interval for \(b\) is \(\WidthExample{}\) on the raw coefficient scale.

\begin{table}[H]
\centering
\caption{Variance Factors and Power for the Four-Wave Worked Example}
\label{tab:worked-example}
\begin{tabular}{crrrrr}
\toprule
Slope & $f_{jj}$ & $s_j$ & Analytical & Monte Carlo & Rejections \\
\midrule
$a$ & 1.057783 & 0.051424 & 99.9946 & 100.00 & 2,997 \\
$b$ & 0.750720 & 0.043322 & 45.4955 & 45.58 & 1,366 \\
$c$ & 0.693712 & 0.041645 & 97.0086 & 96.83 & 2,902 \\
$d$ & 1.061216 & 0.051508 & 99.9944 & 100.00 & 2,997 \\
\bottomrule
\end{tabular}
\par\smallskip{\footnotesize\textit{Note.} Both power columns are percentages. Monte Carlo percentages use the same 2,997 fits for each slope.\par}
\end{table}

\subsection{Population Grid and Matched Model Fitting}

We crossed five population parameter vectors with \(T\in\{3,4\}\) and \(N\in\{200,300,400\}\). Each of the 30 cells used 3,000 independent repetitions, yielding 90,000 simulated fits. Table~\ref{tab:populations} gives the four slopes and the common random intercept variance for each population. The first-wave within-person variances were both one, the first-wave within-person correlation was .30, and the random intercept correlation was .30. We set the innovation covariance to \(\Psi=\Gamma-F\Gamma F^\top\). The analytical and simulation calculations used the same population parameter vectors.

\begin{table}[H]
\centering
\caption{Lagged Coefficients and Random Intercept Variances of the Five Populations}
\label{tab:populations}
\begin{tabular}{rrrrr}
\toprule
$a$ & $b$ & $c$ & $d$ & $\tau_x=\tau_y$ \\
\midrule
0.10 & 0.08 & 0.08 & 0.10 & 3/7 \\
0.10 & 0.16 & 0.08 & 0.30 & 1 \\
0.30 & 0.24 & 0.08 & 0.10 & 1 \\
0.30 & 0.08 & 0.16 & 0.30 & 3/7 \\
0.40 & 0.08 & 0.16 & 0.40 & 3/7 \\
\bottomrule
\end{tabular}
\par\smallskip{\footnotesize\textit{Note.} The two random intercept variances are equal. The within-person and random intercept correlations are each .30.\par}
\end{table}

We used the R package \texttt{powRICLPM} (version 0.2.1) to simulate data from the bivariate RI-CLPM \parencite{Mulder2023powRICLPM}. For each simulated sample, it calls the R package \texttt{lavaan} (version 0.7.2) to fit the model by Gaussian maximum likelihood. The \texttt{within} option constrained the four transition and three innovation covariance parameters across waves. The three initial covariance entries and the three random intercept covariance entries were freely estimated. Measurement error and additional estimation bounds were disabled. For each simulated sample with a usable fit, we tested each of the four unstandardized lagged coefficients separately using a two-sided Wald test of whether the coefficient was zero (\(\alpha=.05\)).

Of the 90,000 simulated fits, 87,807 converged and yielded admissible estimates. For each design and coefficient, we estimated power as the proportion of these usable fits that rejected the null hypothesis.

\subsection{Agreement Across Designs and Slopes}

We first checked the covariance matrix and the standard errors of the four lagged coefficients. For each of the 30 designs, we compared analytical Wald power with simulated power for each of the four lagged coefficients. The mean absolute difference was 0.81 percentage points, and 111 of the 120 comparisons were within two percentage points. As Table~\ref{tab:validation} shows, the mean absolute difference was 1.14 percentage points at three waves and 0.49 at four waves. Every four-wave comparison was within two percentage points in the evaluated grid.

\begin{table}[H]
\centering
\caption{Power Differences by Wave Count and Sample Size}
\label{tab:validation}
\begin{tabular}{rrrrrr}
\toprule
$T$ & $N$ & Fits used (\%) & Mean abs. diff. & Max. abs. diff. & Within 2 points \\
\midrule
3 & 200 & 92.39 & 1.38 & 3.79 & 15/20 \\
3 & 300 & 96.13 & 1.03 & 3.25 & 19/20 \\
3 & 400 & 98.04 & 1.00 & 3.19 & 17/20 \\
4 & 200 & 99.13 & 0.64 & 1.86 & 20/20 \\
4 & 300 & 99.75 & 0.53 & 1.87 & 20/20 \\
4 & 400 & 99.94 & 0.30 & 1.65 & 20/20 \\
\bottomrule
\end{tabular}
\par\smallskip{\footnotesize\textit{Note.} Differences are in percentage points. Each row compares five populations by four slopes. Fits used is the percentage of simulation attempts included in the power comparison.\par}
\end{table}

In the worked example at \(T=4,N=400\), the \(b\) rejection rate was 45.58\%, with a pointwise 95\% Monte Carlo interval of 43.80\%--47.37\%. The analytical value was 45.50\%. The largest discrepancy occurred for \(b\) with three waves, 200 participants, and both autoregressive slopes equal to .40 (Table~\ref{tab:populations}). Analytical power was 13.21\% and simulated power was 9.42\%, a difference of 3.79 percentage points. For this design, 78.20\% of fits entered the comparison. Using all 3,000 attempts as the denominator gives 7.37\%.

We also compared our analytical Wald power results with likelihood-ratio power from the R package \texttt{semPower}, testing the same zero slope under the same model constraints. In regular models, the Wald and likelihood-ratio tests have the same first-order power under local alternatives \parencite{VDV1998}. Across the 120 comparisons, the mean absolute difference between the two power estimates was \LRmean{} percentage points.

\section{Discussion}
\label{sec:discussion}

The choice between the two designs in the introduction depends on which path the study aims to detect. For the population in \eqref{eq:example-population}, allocating 1,200 assessments to 300 people measured four times gives 35.93\% power for the cross-lagged effect \(b\), compared with 22.71\% for 400 people measured three times. The likelihood-ratio analysis in the R package \texttt{semPower} gives power for the chosen hypothesis \parencite{MoshagenBader2024semPower}. Our Wald calculation uses the same standard error to give both power and a planned confidence interval width for each lagged coefficient. Researchers can therefore choose a design that both detects the target effect with adequate power and estimates its size with the desired precision \parencite{Usami2025OptimalDesign}.

Two designs with the same number of assessments can have different costs. In experience sampling studies, \textcite{Revol2025ESMBudget} showed that the choice of \((N,T)\) under a fixed budget depends on how participants are compensated. For dynamic panel models, \textcite{Hecht2024OptimalNT} assigned separate costs to recruiting a participant and measuring that participant at each wave. \textcite{Usami2025OptimalDesign} also allowed a fixed cost per wave when choosing a random intercept panel design to meet a precision target. Given a study's cost function \(C(N,T)\) and budget \(B\), our analytical calculation can quickly screen the pairs satisfying \(C(N,T)\le B\) and identify the one with the highest power for the target lagged effect.

Our power calculations target autoregressive and cross-lagged coefficients. The variance components enter the information matrix because they are estimated jointly with those coefficients. Testing whether a variance component is zero places the null hypothesis on the boundary of the parameter space. In structural equation models, the reference distribution of a likelihood-ratio test can then depend on which constraints hold under the null \parencite{Stoel2006BoundaryConstraints}. In the RI-CLPM specifically, \textcite{SukpanKuiper2026Selecting} compared the chi-square and chi-bar-square difference tests for deciding which random intercepts to include. Extending power analysis to variance components requires specifying the hypothesis and calibrating a test for that boundary.

The derivation assumes that every participant provides both measures at every wave. In practice, a participant may miss a single visit and return, or leave the study before the final wave. Even with the same number of missing assessments, these patterns can leave different amounts of information about a lagged effect. In linear growth models, \textcite{Riordan2020OptimalMissingData} showed how the timing of attrition interacts with the measurement schedule to determine design efficiency. For the RI-CLPM, a first extension would be to calculate information for each observed wave pattern and weight it by its anticipated frequency when missingness is independent of the outcomes. Dropout related to earlier responses would require accounting for how those responses predict continued participation.

The interpretation of simulated power depends on how the analysis handles samples that fail to yield a usable fit. As \textcite{Mulder2023powRICLPM} illustrated, a fitted RI-CLPM can yield an inadmissible estimate such as a negative random intercept variance. Our analytical calculation evaluates information at a specified, regular population point and therefore generates no sample fits to screen. In the three-wave design with the largest discrepancy, the simulated rejection rate for \(b\) was 9.42\% among retained fits, whereas 7.37\% of all simulated studies yielded both a retained fit and a rejection. The first rate gives the rejection probability conditional on a usable fit. The second describes the chance that the planned analysis produces a usable, significant result. How closely analytical power tracks either rate when fitting problems are frequent, and how that comparison changes if investigators refit problematic samples, remain open questions.

Finally, the lagged effects and innovation covariances may change across intervals. \textcite{Mulder2021ThreeExtensions}, for example, examined whether lagged effects could be held equal across waves in an RI-CLPM of sleep and anxiety. The R package \texttt{semPower} also accepts wave-specific lagged effects for likelihood-ratio power calculations \parencite{MoshagenBader2024semPower}. To extend our Wald calculation, one would use the transition matrix \(F_t\) and innovation covariance \(\Psi_t\) for each interval and add one score component per newly freed parameter. For example, freeing the final cross-lagged effect gives it a separate covariance derivative and score, while earlier intervals can retain a common effect.

\section{Conclusion}
\label{sec:conclusion}

To derive analytical Wald power, we first need the partial derivative of the model-implied covariance matrix with respect to each free covariance-structure parameter. We express the RI-CLPM covariance as the sum of a random intercept component and a within-person component that evolves recursively across waves. Differentiating this representation gives each derivative directly and, through the expected Fisher information, yields Wald power for each lagged coefficient. We expect the same approach to extend to RI-CLPMs with time-varying parameters and more than two observed variables by allowing the transition and innovation covariance matrices to vary across intervals and expanding them to include all variables.

Across the designs we evaluated, analytical Wald power was close on average to both the simulated Wald rejection rate and analytical LR power under the same model constraints. Agreement with simulated Wald rejection rates was closest in the four-wave designs. By avoiding repeated model fitting, the analytical method saves substantial time when comparing power and confidence interval widths across many combinations of sample size \(N\) and number of waves \(T\).

\printbibliography
\end{document}